\documentclass[conference]{IEEEtran}

\usepackage{multicol}
\usepackage{amsmath,amssymb,cite,multirow,subfigure}
\usepackage{graphicx}
\usepackage{url}
\usepackage[linesnumbered,ruled, vlined]{algorithm2e}
\usepackage{algorithmic}
\usepackage{makecell}

\usepackage{booktabs}
\usepackage{threeparttable}

\begin{document}
%
\title{Reduced Complexity Belief Propagation Decoders \\for Polar Codes}

\author{\authorblockN{Jun Lin, Chenrong Xiong and Zhiyuan Yan}
\authorblockA{Department of Electrical and Computer Engineering, Lehigh University, PA, USA\\
Email: \{jul311,chx310,yan\}@lehigh.edu}}

\maketitle

\begin{abstract}
Polar codes are newly discovered capacity-achieving codes, which have attracted lots of research efforts. Polar codes can be efficiently decoded by the low-complexity successive cancelation (SC) algorithm and the SC list (SCL) decoding algorithm. The belief propagation (BP) decoding algorithm not only is an alternative to the SC and SCL decoders, but also provides soft outputs that are necessary for joint detection and decoding. Both the BP decoder and the soft cancelation (SCAN) decoder were proposed for polar codes to output soft information about the coded bits. In this paper, first a belief propagation decoding algorithm, called reduced complexity soft cancelation (RCSC) decoding algorithm, is proposed. Let $N$ denote the block length. Our RCSC decoding algorithm needs to store only $5N-3$ log-likelihood ratios (LLRs), significantly less than $4N-2+\frac{N\log_2N}{2}$ and $N(\log_2N+1)$ LLRs needed by the BP and SCAN decoders, respectively, when $N\geqslant 64$. Besides, compared to the SCAN decoding algorithm, our RCSC decoding algorithm eliminates unnecessary additions over the real field. Then the simplified SC (SSC) principle is applied to our RCSC decoding algorithm, and the resulting SSC-aided RCSC (S-RCSC) decoding algorithm further reduces the computational complexity. Finally, based on the S-RCSC decoding algorithm, we propose a corresponding memory efficient decoder architecture, which has better error performance than existing architectures. Besides, our decoder architecture consumes less energy on updating LLRs.

\end{abstract}

\section{Introduction}
\label{sec:intro}
Polar codes~\cite{arikan} are a significant breakthrough in coding theory, since polar codes can achieve the channel capacity of binary-input symmetric memoryless channels~\cite{arikan} and arbitrary discrete memoryless channels~\cite{sas_polar}. Polar codes of block length $N$ can be efficiently decoded by a successive cancelation (SC) algorithm~\cite{arikan} with a complexity of $O(N\log N)$, where $N$ is the block length. In spite of the low-complexity nature of the SC algorithm, the error performance of the SC algorithm is worse than Turbo or LDPC codes for short or moderate polar codes.


The belief propagation (BP) decoding of polar codes over factor graph~\cite{factor_graph} was proposed in~\cite{arikan}. The message passing schedules and error performances under finite lengths were further discussed in~\cite{bp_polar_org,finite_length_polar}. A low complexity soft-output version of the SC decoder called soft cancelation (SCAN) decoder was proposed in~\cite{soft_sc_bp}. Compared to the BP decoders in~\cite{bp_polar_org,finite_length_polar}, the SCAN decoder has much lower computational complexity and requires less memory to store the soft messages. The SCAN decoding algorithm employs a serial message updating schedule, which is similar to the SC decoding of polar codes. In contrast, the BP decoding algorithms in~\cite{arikan,bp_polar_org,finite_length_polar} employ a parallel message updating schedule. It was shown in~\cite{soft_sc_bp} that the SCAN decoding algorithm converges faster than the BP decoding algorithm due to the better dissemination of information. Compared to the SC decoding algorithm, both the BP and SCAN decoding algorithms have higher computational complexity and require more memory. However, these belief propagation decoding algorithms not only offer an alternative to the SC and SC list (SCL) decoders, but also are necessary for the application of polar codes in receivers that employ the joint detection and decoding technique.

Even though the decoder architectures of the SC and SCL decoding algorithms have been well studied, the architectures of the soft-output decoders for polar codes have not been sufficiently investigated in literature. Under a 65nm CMOS technology, for a (1024, 512) polar code, the BP polar decoder chip in~\cite{bp_polar_chip_umich} achieves a coded throughput of 4.68Gbps when the signal to noise ratio (SNR) equals 4.0dB by occupying 1.48mm$^2$ silicon area. Several early stopping criterions for the BP decoding of polar codes were proposed in~\cite{yuan_bp_tsp} to improve the decoder throughput and energy efficiency. The (1024, 512) polar BP decoder in~\cite{yuan_bp_tsp} achieves a net information throughput of 4.5 Gbps using 1.96 million gates under a 45nm CMOS technology. The BP decoder architectures in~\cite{bp_polar_chip_umich,yuan_bp_tsp} achieve multi Gbps throughput due to the parallel message updating schedule. However, these architectures in~\cite{bp_polar_chip_umich,yuan_bp_tsp} are not suitable for larger block lengths, since the number of basic processing elements (PE) is $N$ and $\frac{N}{2}\log_2N$, respectively. For larger $N$, the resulting BP decoders will suffer from excessive area and power. For the SCAN decoding algorithm, its efficient hardware implementations have not been discussed in open literature to the best of our knowledge.

In this paper,  we focus on the soft-output decoding of polar codes. Our main contributions are:

(1)  A reduced complexity soft-output version of SC decoder, called \emph{reduced complexity soft cancelation} (RCSC) decoder, is proposed. Compared to the BP and SCAN decoding algorithms, our RCSC decoding algorithm has lower computational complexity and stores less LLRs. Our RCSC decoding algorithm needs to store only $5N-3$ LLRs, significantly less than $4N-2+\frac{N\log_2N}{2}$ and $N(\log_2N+1)$ LLRs needed by the BP and SCAN decoder, respectively, when $N\geqslant 64$. Besides, our RCSC decoding algorithm converges almost as fast as the SCAN decoding algorithm.

(2) The simplified SC~\cite{low_latency_polar} (SSC) principle is applied to our RCSC decoding algorithm, resulting in the SSC-aided RCSC (S-RCSC) decoding algorithm with even less computational complexity.

(3) Based on our S-RCSC algorithm, a corresponding scalable decoder architecture for polar codes is proposed. Compared to BP decoders in~\cite{bp_polar_chip_umich,yuan_bp_tsp}, our decoder architecture has better error performance. Besides, our decoder architecture consumes less energy on updating LLRs.


\section{Preliminaries}\label{sec: pre}

\subsection{Polar codes} \label{ssec:polar_encoding}
Under the Ar{\i}kan's construction method~\cite{arikan}, the generation matrix of a polar code is an $N\times N$ matrix $G=B_NF^{\otimes n}$, where $N=2^n$, $B_N$ is the bit reversal permutation matrix~\cite{arikan}, and $F=\left[{1\atop 1}{0\atop 1}\right]$. Here $\otimes n$ denotes the $n$th Kronecker power and $F^{\otimes n} = F\otimes F^{\otimes (n-1)}$. Let $u_0^{N-1} = (u_0,u_1,\cdots,u_{N-1})$ denote the data bit sequence and $x_0^{N-1} = (x_0,x_1,\cdots,x_{N-1})$ the coded bit sequence, then $x_0^{N-1}=u_0^{N-1}G$. An $(N,K)$ polar code is defined by setting $N-K$ bits of $u_0^{N-1}$ to zero, denoted as frozen bits. The information is conveyed by the rest $K$ bits, denoted as information bits. 

\subsection{BP and SCAN Decoding Algorithms for Polar Codes} \label{ssec:bp_scan}
\begin{figure} [hbt]
\centering
  \includegraphics[width=3in]{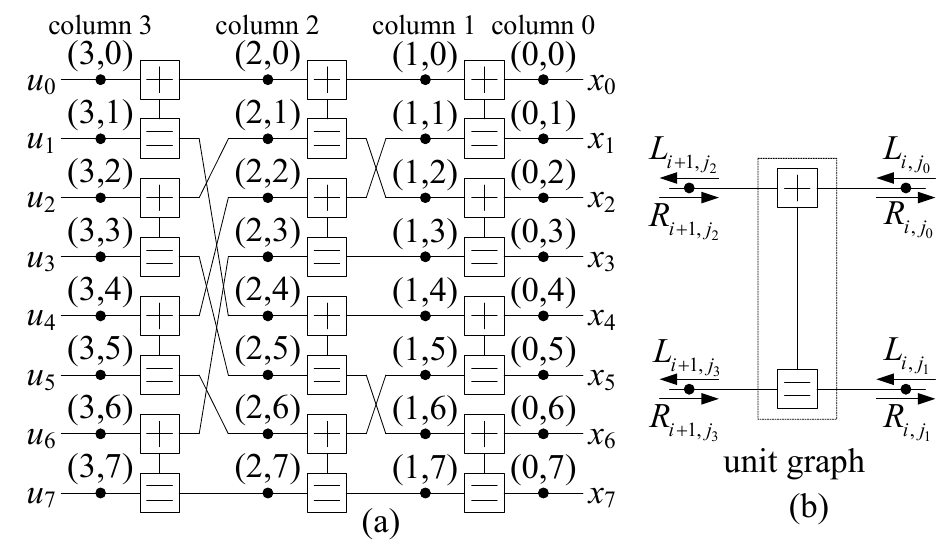}
  \caption{(a) Factor graph for a polar code with $N=8$ (b) Message passing on an unit graph}\label{fig: factor_graph}
\end{figure}

The BP and SCAN decoding algorithms are performed over the factor graph of a polar code, which is derived from the encoding equations. Take $N=8$ as an example, the corresponding factor graph is shown in Fig.~\ref{fig: factor_graph}(a). For $i=0,1,2$ and $j=0,1,\cdots,7$, node $(i,j)$ in Fig.~\ref{fig: factor_graph}(a) has two associated LLR messages $L_{i,j}$ and $R_{i,j}$, which are passed to the left and right directions of the factor graph. $L_{i,j}$ and $R_{i,j}$ are called left and right LLR messages, respectively. As shown in Fig.~\ref{fig: factor_graph}(a), the factor graph of a polar code of length $N=2^n$ consists of $\frac{Nn}{2}$ homogenous unit factor graphs and $n+1$ columns of nodes.

The message passing on a unit graph is shown in Fig.~\ref{fig: factor_graph}(b), where $L_{i,j_0},L_{i,j_1},R_{i+1,j_2},R_{i+1,j_3}$ are LLR messages sending to the unit graph. $R_{i,j_0},R_{i,j_1},L_{i+1,j_2},L_{i+1,j_3}$ are LLR messages sending from the unit factor graph and can be computed as follows:
\begin{eqnarray}
&& L_{i+1,j_2} = f(R_{i+1,j_3}+L_{i,j_1}, L_{i,j_0}), \label{eqn: updating_rule1}\\
&& L_{i+1,j_3} = f(R_{i+1,j_2},L_{i,j_0})+L_{i,j_1}, \label{eqn: updating_rule2}\\
&& R_{i,j_0} = f(R_{i+1,j_2}, R_{i+1,j_3} + L_{i,j_1}), \label{eqn: updating_rule3}\\
&& R_{i,j_1} = R_{i+1,j_3} + f(R_{i+1,j_2}, L_{i,j_0}), \label{eqn: updating_rule4}
\end{eqnarray}
where $f(a,b)$ is approximated as
\begin{equation}
\label{equ: f_app}
f(a,b)\approx \mbox{sign}(a)\times\mbox{sign}(b)\times\min(|a|,|b|).
\end{equation}


For $i=0,1,\cdots,n$, let $\mathbf{L}_i = (L_{i,0},L_{i,1},\cdots,L_{i,N-1})$ and $\mathbf{R}_i = (R_{i,0},R_{i,1},\cdots,R_{i,N-1})$ denote all the left and right LLRs associated with the nodes in column $i$, respectively. Each iteration of the BP decoding algorithm consists of the left-direction message updating and the following right-direction message updating. During the left-direction message updating, $\mathbf{L}_1, \mathbf{L}_2,\cdots,\mathbf{L}_n$ are updated in serial using Eqs.~(\ref{eqn: updating_rule1}) and~(\ref{eqn: updating_rule2}), where all the left LLRs associated with the same column are updated in parallel. During the right-direction message updating, $\mathbf{R}_{n-1}, \mathbf{R}_{n-2},\cdots,\mathbf{R}_0$ are updated in serial using Eqs.~(\ref{eqn: updating_rule3}) and~(\ref{eqn: updating_rule4}), where all the right LLRs associated with the same column are computed in parallel. Note that $\mathbf{L}_0$ is the received channel LLR vector and $\mathbf{R}_n$ has $N$ constant LLRs, where $R_{n,j}=0$ if $u_i$ is an information bit and $R_{n,j}=+\infty$ otherwise. As shown in~\cite{bp_polar_chip_umich}, a BP decoder needs to store $N(\log_2N+1)$ LLRs. In terms of the computational complexity, each iteration of the BP decoding algorithm is dominated by $2N\log_2N$ additions and $2N\log_2N$ comparisons over the real field.

Based on Eqs.~(\ref{eqn: updating_rule1}) to~(\ref{eqn: updating_rule4}), the SCAN decoding algorithm applies a different message updating schedule, which follows the SC decoding schedule. For each $\mathbf{L}_i$ and $\mathbf{R}_i$, instead of updating all the LLRs in parallel, the SCAN decoding algorithm divides $N$ LLRs into $2^i$ groups, which are updated in serial. However, the $\frac{N}{2^i}$ LLRs within each group are updated in parallel. Compared to the BP decoding algorithm, the SCAN decoding algorithm converges faster and needs to store fewer LLRs due to the fact that certain LLRs will never be used in the following iterations. As shown in~\cite{soft_sc_bp}, the SCAN decoding algorithm needs to store $2N-1$ left LLRs. In~\cite{soft_sc_bp}, the right LLRs are divided into two groups: $R_{i,j}$'s with $j$ being an odd and even number, which are denoted as $\mathbf{R}^o$ and $\mathbf{R}^e$, respectively. The SCAN decoding algorithm in~\cite{soft_sc_bp} needs to store $\frac{N}{2}\log_2N$ and $2N-1$ LLRs for $\mathbf{R}^o$ and $\mathbf{R}^e$, respectively. Note that, the SCAN decoding algorithm has the same computational complexity per iteration as the BP decoding algorithm.

\section{The Proposed RCSC Decoding Algorithm} \label{sec: rcsc}
\subsection{Modified Message Passing on the Unit Graph}\label{ssec: modified_msg_passing}

\begin{figure} [hbt]
\centering
  \includegraphics[width=2.4in]{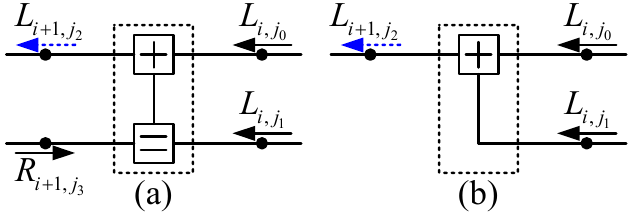}
  \caption{(a) Message passing for the computing of $L_{i+1,j_2}$ when $i=0$ (b) Message passing for the computing of $L_{i+1,j_2}$ when $i>0$} \label{fig: rcsc_unit_graph}
\end{figure}

Both the BP and SCAN decoding algorithms are based on the message passing schedules shown in Eqs.~(\ref{eqn: updating_rule1}) to~(\ref{eqn: updating_rule4}). In this section, we first propose a modified message passing schedule to compute $\mathbf{L}^e$, which are $L_{i,j}$'s with $j$ being an even integer. For each unit graph, the left LLR
\begin{equation}
L_{i+1,j_2} = \left\{ \begin{array}{ll}
 f(R_{i+1,j_3}+L_{i,j_1}, L_{i,j_0}) &\mbox{ if $i=0$}, \\
 f(L_{i,j_1}, L_{i,j_0}) &\mbox{ if $i>0$},
       \end{array} \right.
\end{equation}
where $j_2$ is an even integer and $j_3=j_2+1$. The modified message passing schedule for $i=0$ and $i>0$ is shown in Figs.~\ref{fig: rcsc_unit_graph}(a) and \ref{fig: rcsc_unit_graph}(b), respectively. As shown in Figs.~\ref{fig: rcsc_unit_graph}(a) and \ref{fig: rcsc_unit_graph}(b), the right LLR $R_{i+1,j_3}$ is considered only when $i=0$. For the computation of $ L_{i+1,j_3}$ and other right LLRs, Eqs.~(\ref{eqn: updating_rule2}) to~(\ref{eqn: updating_rule4}) are still used.

Compared to the message passing schedule shown in Fig.~\ref{fig: factor_graph}(b), the benefits of our modified message passing schedule are as follows.

1) When $i>1$, an summation over the real field is saved when computing $L_{i,j}$ with $j$ being an even integer.

2) When $j$ is an even integer, the data dependency between $R_{i,j+1}$ and $L_{i,j}$ is removed when $i>1$. As a result, it is unnecessary to store all LLRs in $\mathbf{R}^o$ for the following iterations.

When $i=0$, the modified message passing schedule still considers the right LLR as $R_{i+1,j_3}$ as shown in Fig.~\ref{fig: rcsc_unit_graph}(a). The reason is that we want to find a way to feed $\mathbf{R}^o$ back to the next iteration. 

\subsection{The Proposed RCSC Decoding Algorithm}
\label{ssec: rcsc}
Based on our modified message passing schedule, an RCSC decoding algorithm is proposed in Alg.~\ref{algo: rcsc} for polar codes. Like the SCAN decoding algorithm, our RCSC algorithm also employs the SC decoding schedule when updating all soft messages. As shown in Alg.~\ref{algo: rcsc}, $\mathbb{L}$ is a set of $n+1$ LLR arrays $\mathcal{L}_0, \mathcal{L}_1,\cdots,\mathcal{L}_{n}$, where $\mathcal{L}_i$ stores $2^{n-i}$ LLRs for $i=0,1,\cdots,n$. $\mathbb{R}^o$ is a set of $n$ LLR arrays $\mathcal{R}^o_1, \mathcal{R}^o_2,\cdots,\mathcal{R}^o_{n}$, where $\mathcal{R}^o_i$ stores $2^{n-i}$ LLRs for $i=1,2,\cdots,n$. $\mathbb{R}^e$ is a set of $n+1$ LLR arrays $\mathcal{R}^e_0, \mathcal{R}^e_1,\cdots,\mathcal{R}^e_{n}$, where $\mathcal{R}^e_i$ stores $2^{n-i}$ LLRs for $i=0,1,\cdots,n$. For our RCSC algorithm, $\mathbb{L}, \mathbb{R}^o, \mathbb{R}^e$ are the memory locations used to store left LLRs, $\mathbf{R}^o$ and $\mathbf{R}^e$, respectively.

\begin{algorithm}
\caption{The Proposed RCSC Decoding algorithm}
\label{algo: rcsc}
\DontPrintSemicolon
\SetKwInOut{Input}{input}\SetKwInOut{Output}{output}

\Input{$n, \mbox{the received channel message } y_0^{N-1}$}
\Output{$\hat{x}_0^{N-1}$}
\BlankLine
\For{$\mbox{iter}=1$ \KwTo $I_M$} {
\For{$j =0$ \KwTo $N-1$}{
$\mbox{LComp}(j, \mathbb{L},\mathbb{R}^o, \mathbb{R}^e)$\;
\If{$j\mod 2=0$}{
\lIf{$u_j$ is a frozen bit}{$\mathcal{R}^e_n[0] = \infty$}
\lElse{$\mathcal{R}^e_n[0] = 0$}
}\Else {
\lIf{$u_j$ is a frozen bit}{$\mathcal{R}^o_n[0] = \infty$}
\lElse{$\mathcal{R}^o_n[0] = 0$}
$\mbox{RComp}(j, \mathbb{L},\mathbb{R}^o, \mathbb{R}^e)$\;
}
}

\For{$j =0$ \KwTo $N-1$}{
\lIf{$(\mathcal{L}_0[j]+\mathcal{R}^e_0[j]) \geqslant 0$} {$\hat{x}_j = 0$} \lElse{$\hat{x}_j = 1$}
}

\lIf{$\hat{x}_0^{N-1}$ is a valid codeword} {return $\hat{x}_0^{N-1}$}
}
\end{algorithm}

\begin{algorithm}
\caption{LComp$(j, \mathbb{L},\mathbb{R}^o, \mathbb{R}^e)$}
\label{algo: left_comp}
\DontPrintSemicolon
\SetKwInOut{Input}{input}\SetKwInOut{Output}{output}

\Input{$j, \mathbb{L},\mathbb{R}^o, \mathbb{R}^e$}
\BlankLine

$i=s_j$\;
\For{$k = 0$ \KwTo $2^{n-i}-1$} {
\If{$j=0$}{
$\mathcal{L}_i[k] = f(\mathcal{L}_{i-1}[2k], \mathcal{L}_{i-1}[2k+1]+\mathcal{R}^o_i[k])$\;
} \lElse {
$\mathcal{L}_i[k] = \mathcal{L}_{i-1}[2k+1] + f(\mathcal{L}_{i-1}[2k], \mathcal{R}^e_i[k])$
}
}

\For{$i = s_j+1$ \KwTo $n$} {
\For{$k =0$ \KwTo $2^{n-i}-1$}{
$\mathcal{L}_i[k]=f(\mathcal{L}_{i-1}[2k],\mathcal{L}_{i-1}[2k+1])$\;
}
}
\end{algorithm}

\begin{algorithm}
\caption{RComp$(j, \mathbb{L},\mathbb{R}^o, \mathbb{R}^e)$}
\label{algo: right_comp}
\DontPrintSemicolon
\SetKwInOut{Input}{input}\SetKwInOut{Output}{output}

\Input{$j, \mathbb{L},\mathbb{R}^o, \mathbb{R}^e$}
\BlankLine

\For{$i = n-1$ \KwTo $e_j+1$} {
\For{$k =0$ \KwTo $2^{n-i}-1$}{
$\mathcal{R}^o_i[2k]=f(\mathcal{R}^e_{i+1}[k], \mathcal{R}^o_{i+1}[k] + \mathcal{L}_{i}[2k+1])$\;
$\mathcal{R}^o_i[2k+1]=\mathcal{R}^o_{i+1}[k] + f(\mathcal{R}^e_{i+1}[k], \mathcal{L}_{i}[2k])$\;
}
}
$i=e_j$\;
\For{$k = 0$ \KwTo $2^{n-i}-1$} {
$\mathcal{R}^e_i[2k]=f(\mathcal{R}^e_{i+1}[k], \mathcal{R}^o_{i+1}[k] + \mathcal{L}_{i}[2k+1])$\;
$\mathcal{R}^e_i[2k+1]=\mathcal{R}^o_{i+1}[k] + f(\mathcal{R}^e_{i+1}[k], \mathcal{L}_{i}[2k])$\;
}

\end{algorithm}

As shown in Alg.~\ref{algo: rcsc}, each iteration of our RCSC decoding algorithm is divided into $N$ serial step. For $i=0,1,\cdots,N-1$, during step $i$, part of left LLRs are first updated in the way shown in Alg.~\ref{algo: left_comp}. If $i$ is an odd integer, part of the right LLRs are also updated in the way shown in Alg.~\ref{algo: right_comp}. $s_j$ and $e_j$ in Alg.~\ref{algo: left_comp} and~\ref{algo: right_comp}, respectively, are variable integer indices depending on $j$. Let $(b_{n-1}, b_{n-2},\cdots,b_0)$ be the binary representation of the integer index $j$, where $b_0$ is the least significant bit (LSB). When $j$ is a non-zero even number, $s_j = n-(k+1)$ such that $b_{k+1}=1$ and $b_{r=0}$ for $r\leqslant k$. When $j=0$, $s_j=1$. When $b_0=1$, $s_j=n$. $e_j = n-k$, where $k$ is the smallest integer such that $b_k =0$. If $b_k=1$ for $k=0,1,\cdots,n-1$, then $e_j=0$.

\begin{table}[!hbt]
  \centering
  \caption{Comparisons with the BP and SCAN Decoding Algorithms}
  \label{tab:complexity_cmp}
  \footnotesize
  \begin{tabular}{c|c|c||c}
    \hline
      & BP & SCAN & RCSC \\ \hline\hline
     \# of stored LLRs & $N(n+1)$ & $4N-2+\frac{Nn}{2}$ & $5N-3$ \\ \hline
     \# of additions & $2Nn$ & $2Nn$ & $\frac{3Nn}{2}+\frac{N}{2}$ \\ \hline
     \# of comparisons & $2Nn$ & $2Nn$ & $2Nn$ \\ \hline
  \end{tabular}
\end{table}
In Table~\ref{tab:complexity_cmp}, we compare our RCSC decoding algorithm with the BP and SCAN decoding algorithms in terms of memory and computational complexities. The block length $N=2^n$. As shown in Table~\ref{tab:complexity_cmp}, when $n\geqslant 6\mbox{ }(N\geqslant 64)$, the number of stored LLRs by our RCSC algorithm is the smallest among all three algorithms. When the block lengths are larger, our RCSC algorithm achieves more significant memory saving compared to the BP and SCAN decoding algorithms. Besides, as shown in Table~\ref{tab:complexity_cmp}, our RCSC algorithm saves $\frac{(n-1)N}{2}$ additions per iteration compared to the BP and SCAN algorithm.

Compared to the SCAN decoding algorithm, the major improvements of our RCSC decoding algorithm are as follows.

1) As shown in Alg.~\ref{algo: left_comp}, when updating left LLRs, our RCSC algorithm employs the modified message passing schedule shown in Section~\ref{ssec: modified_msg_passing}, which changes the data dependency between left LLRs and $\mathbf{R}^o$.

2) The modified data dependency results in efficient storage of $\mathbf{R}^o$ and reduced number of additions.

\subsection{SSC-aid RCSC Decoding Algorithm}\label{ssec: s-rcsc}
\begin{figure} [hbt]
\centering
  \includegraphics[width=2.8in]{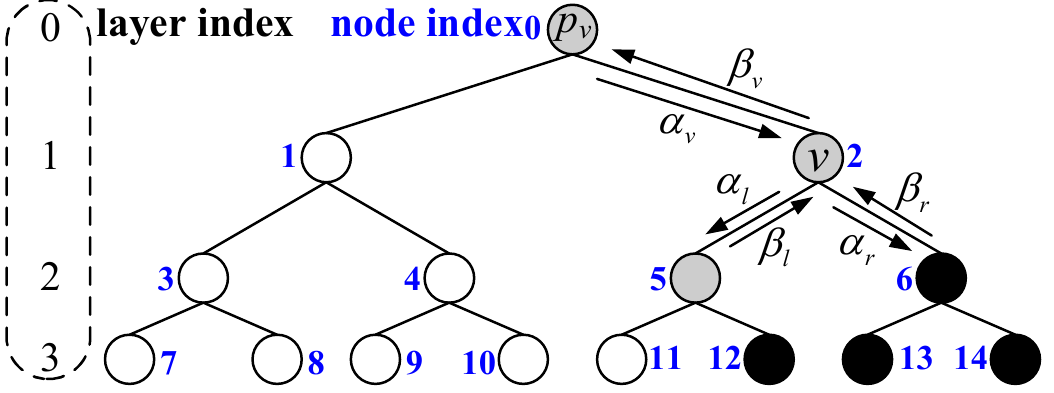}
  \caption{Message passing on a binary tree}\label{fig: dec_tree}
\end{figure}

In~\cite{low_latency_polar, fast_polar_SC_gross}, a polar code of length $N$ can be represented by a binary tree of depth $n$, where each node represents a constituent code. Fig.~\ref{fig: dec_tree} shows the tree representation of an (8, 3) polar code, where the black and white leaf nodes correspond to information and frozen bits, respectively. Among all the nodes of a tree, a node is called a rate-1 and rate-0 node if all of its leaves are associated with information and frozen bits, respectively. The SC decoding algorithm was performed on a binary tree in~\cite{low_latency_polar, fast_polar_SC_gross}, where each node behaves as a decoder for the corresponding constituent code. In this section, we formulate the proposed RCSC decoding algorithm on a binary tree.

As shown in Fig.~\ref{fig: dec_tree}, a node $v$ with layer index $i$ and a constituent code of length $\mathcal{N}_v=2^{n-i}$
receives a soft message vector, $\alpha_v$, containing $\mathcal{N}_v$ LLRs, from its parent
node $v_p$. Node $v$ then calculates the soft message vector to its left child, $\alpha_l$,
containing $\mathcal{N}_v/2$ LLRs, via
\begin{equation}
\label{eqn: left_vector_comp}
\alpha_l[k] = \left\{ \begin{array}{ll}
 f(\beta_r[k]+\alpha_v[2k+1], \alpha_v[2k]) &\mbox{ if $i=0$}, \\
 f(\alpha_v[2k], \alpha_v[2k+1]) &\mbox{ if $i>0$},
       \end{array} \right.
\end{equation}
for $k=0,1,\cdots,\mathcal{N}_v/2-1$, where $\beta_r$ (containing $\mathcal{N}_v/2$ LLRs) is the soft message vector sent from the right child in the previous iteration. Node $v$ then waits until an updated soft LLR vector, $\beta_l$ (containing $\mathcal{N}_v/2$ LLRs), is sent from its left child. In the following step, Node $v$ calculates the soft message vector to its right child, $\alpha_r$, where
$\alpha_r[k]$ = $\alpha_v[2k+1] + f(\alpha_v[2k], \beta_l[k])$
for $k=0,1,\cdots,\mathcal{N}_v/2-1$. Once both $\beta_l$ and $\beta_r$ are updated, the soft message sent from node $v$, $\beta_v$, is calculated using
$\beta_v[2k]$ = $f(\beta_l[k], \beta_r[k] + \alpha_v[2k+1])$ and
$\beta_v[2k+1]$ = $\beta_r[k] + f(\beta_l[k], \alpha_v[2k])$
for $k=0,1,\cdots,\mathcal{N}_v/2-1$. If node $v$ is a leaf node, $\beta_v =0$ if node $v$ corresponds to an information bit and $\beta_v=+\infty$ otherwise.

For our SSC-aid RCSC (S-RCSC) decoding algorithm, if node $v$ is a rate-0 node, $\beta_v=(+\infty, +\infty, \cdots, +\infty)$ is returned immediately without traversing its child nodes. If node $v$ is a rate-1 node, $\beta_v=(0, 0, \cdots, 0)$ is returned immediately. In this way, both the decoding latency and computational complexity can be reduced further. The S-RCSC algorithm has the same memory architecture as that of RCSC. $\beta_v$ is stored in $\mathbb{R}^e$ and $\mathbb{R}^o$ if node $v$ is the left and right child of its parent node $v_p$, respectively. For each node $v$, $\alpha_v$ can be stored in $\mathbb{L}$. Take the (8, 3) polar code in Fig.~\ref{fig: dec_tree} as an example, during each iteration, the RCSC decoding algorithm needs to visit all 15 nodes. In contrast, our S-RCSC decoding algorithm visits only 7 nodes (nodes 0, 1, 2, 5, 6, 11, 12) in each iteration. Hence, the S-RCSC decoding algorithm has lower computational complexity that the RCSC decoding algorithm since the number of LLRs that need to be updated is reduced due to the SSC principle.

\subsection{Numerical Results}\label{ssec: sim_results}
\begin{figure} [hbt]
\centering
  \includegraphics[width=2.2in]{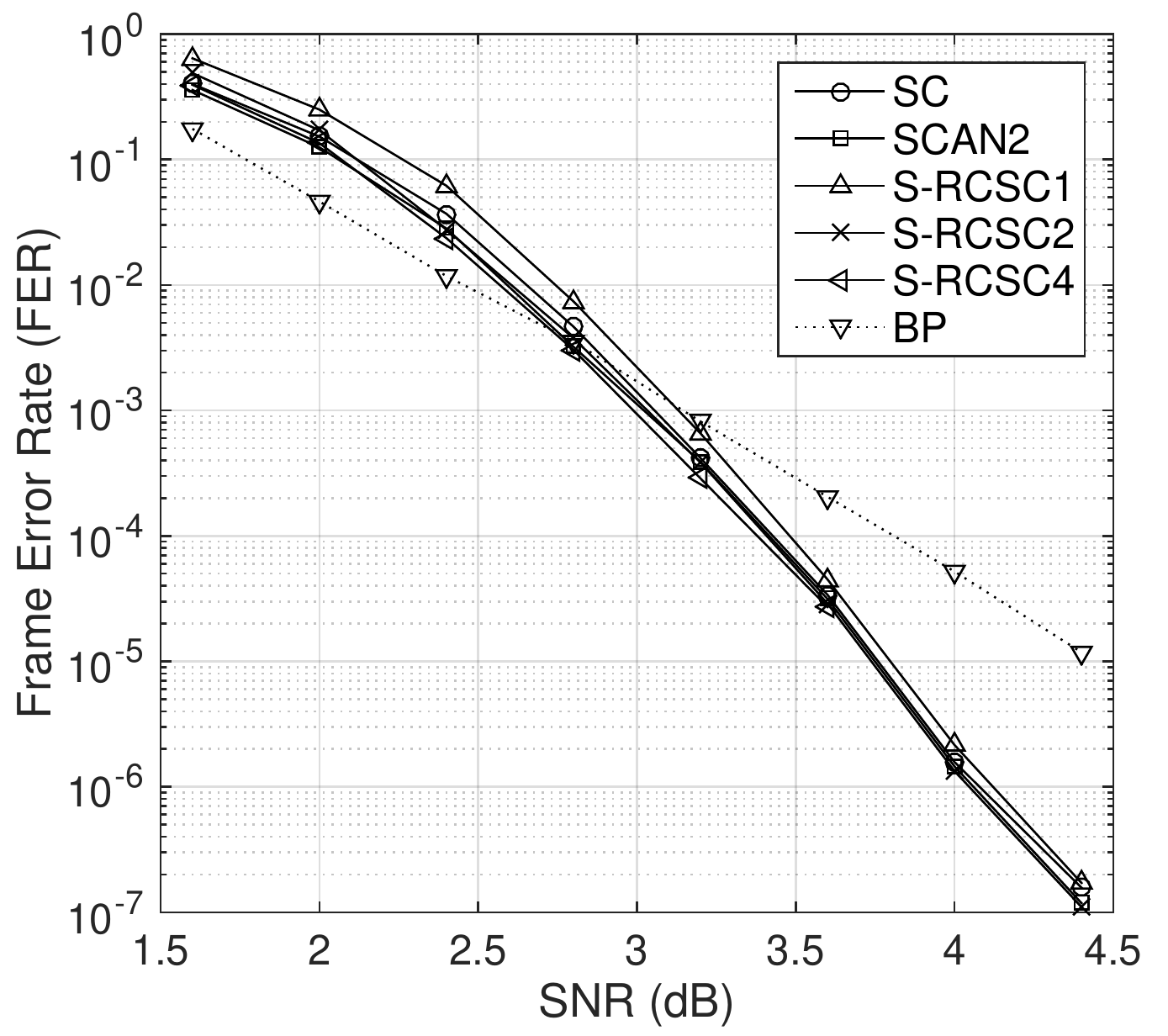}
  \caption{FER performances of a (1024, 512) polar code}\label{fig: fer1024}
\end{figure}

\begin{figure} [hbt]
\centering
  \includegraphics[width=2.2in]{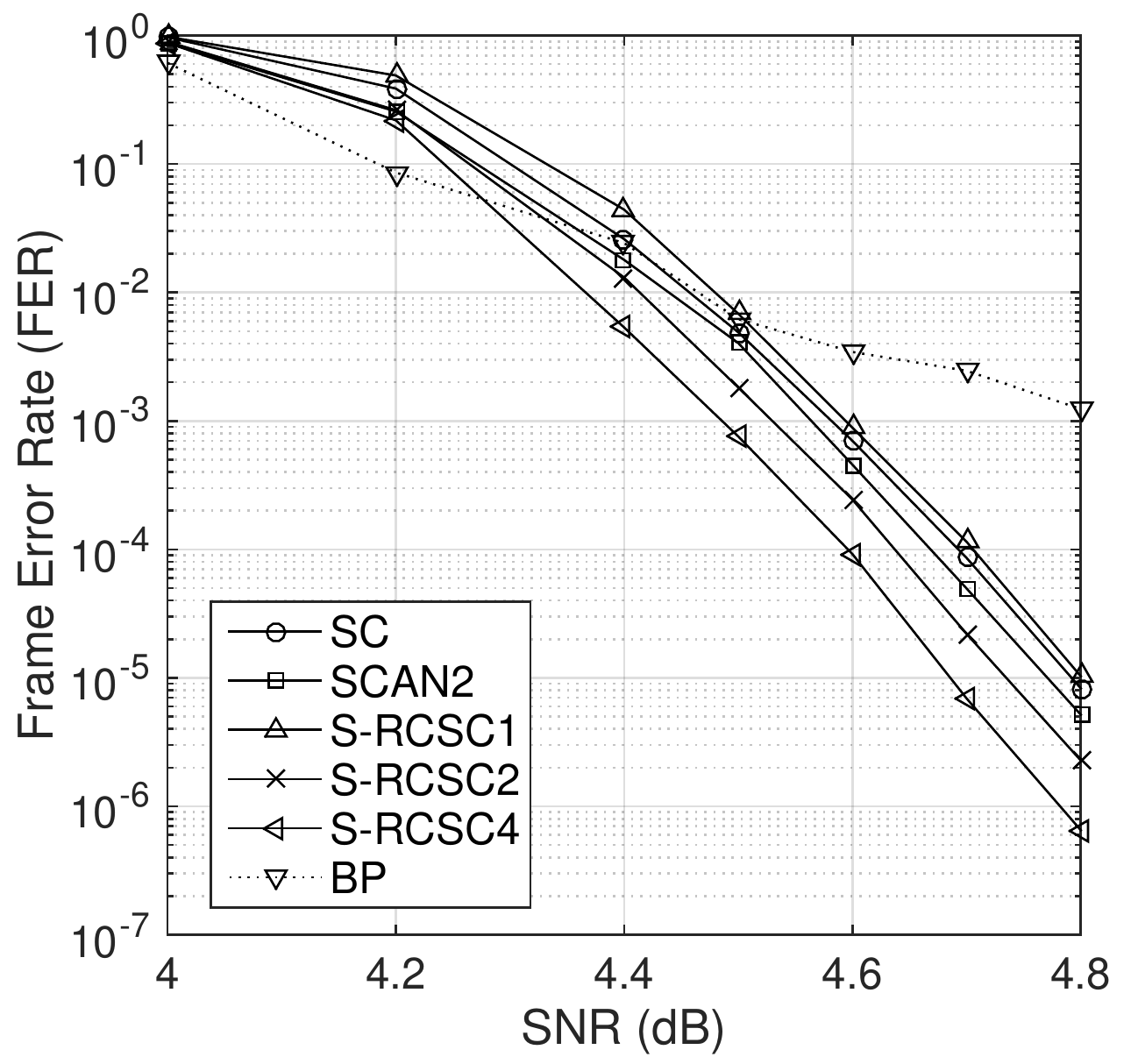}
  \caption{FER performances of a (32768, 29504) polar code}\label{fig: fer32k_r0_9}
\end{figure}

For both a (1024, 512) and a (32768, 29504) polar codes, the frame error rate (FER) performance of our S-RCSC decoding algorithm is shown in Figs.~\ref{fig: fer1024} and~\ref{fig: fer32k_r0_9}, respectively, where S-RCSC$k$ denotes our S-RCSC algorithm with the maximum number of iterations $I_M=i$. SCAN2 denotes the SCAN decoding algorithm with the maximum number of iterations being 2. The error performances of the BP decoding algorithm under 100 maximum iterations are also shown in Figs.~\ref{fig: fer1024} and~\ref{fig: fer32k_r0_9}. As shown in Figs.~\ref{fig: fer1024} and~\ref{fig: fer32k_r0_9}, the error performance of the S-RCSC is slightly better than SCAN decoding algorithm. For both polar codes, our S-RCSC decoding algorithm has better error performance than the BP decoding algorithm when the signal to noise ratio (SNR) is higher. Note that the error performances of the RCSC and S-RCSC decoding algorithm are the same. As a result, we did not show the error performances of the RCSC decoding algorithm in Figs.~\ref{fig: fer1024} and~\ref{fig: fer32k_r0_9}. Let $I_{av}$ and $I_{av}'$ denote the average numbers of iterations for the S-RCSC and SCAN decoding algorithms, respectively, when the maximum number of iterations is set to two. As shown in Table~\ref{tab:av_iter_floating}, the S-RCSC algorithm and converges almost as fast as the SCAN algorithm.

\begin{table}[!hbt]
  \centering
  \caption{The Average Number of Iterations When $I_M=2$}
  \label{tab:av_iter_floating}
  \footnotesize
  \begin{tabular}{c|c|c|c|c|c|c}
    \hline
    code &\multicolumn{6}{c}{$(1024, 512)$} \\ \hline
     SNR & 1.6 & 2.0 & 2.4 & 2.8 & 4.0 &4.4  \\ \hline
     $I_{av}$ & 1.62 & 1.25 & 1.06 & 1.007 & 1.000002 & 1 \\ \hline
      $I_{av}'$ & 1.36 & 1.12 & 1.02 & 1.003 & 1.000002 & 1 \\ \hline\hline
     code & \multicolumn{6}{c}{$(32768, 29504)$} \\ \hline
     SNR & 4.0 & 4.2 & 4.5 & 4.6 & 4.7 &4.8  \\ \hline
     $I_{av}$ & 1.97 & 1.41 & 1.006 & 1.0009 & 1.0001 & 1.00001 \\ \hline
     $I_{av}'$ & 1.87 & 1.25 & 1.004 & 1.0004 & 1.00005 & 1.000005 \\ \hline\hline
  \end{tabular}
\end{table}

\section{An Efficient Soft-Output Decoder Architecture for Polar Codes} \label{sec: hw_archi}
\subsection{Top Architecture}
\begin{figure} [hbt]
\centering
  \includegraphics[width=3in]{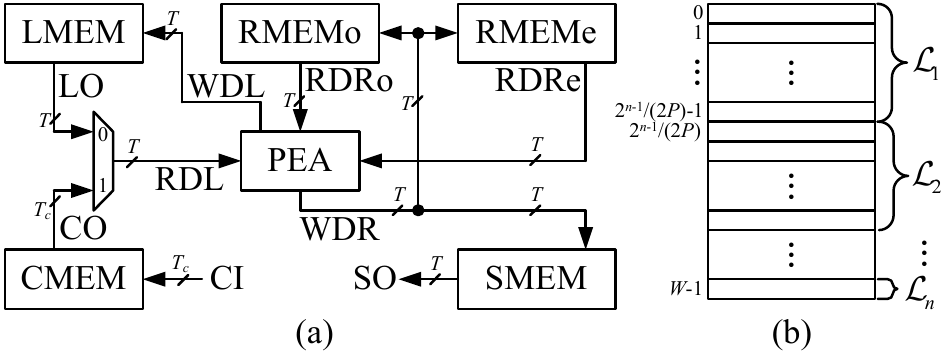}
  \caption{(a) Decoder top architecture (b) LLR arrangement in LMEM}\label{fig: top_archi}
\end{figure}

\begin{figure} [hbt]
\centering
  \includegraphics[width=3in]{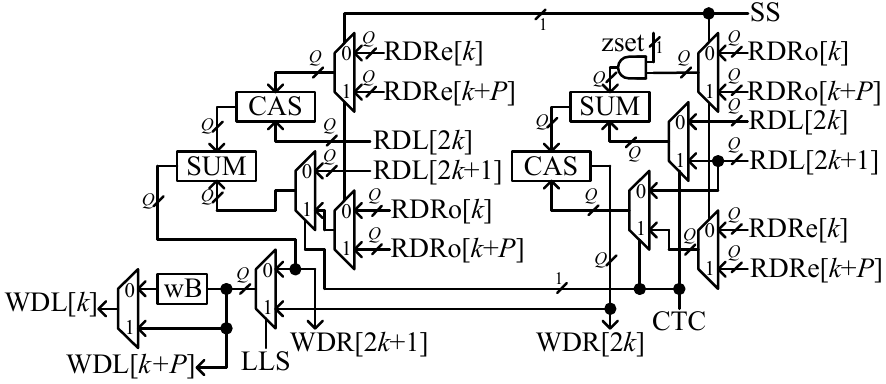}
  \caption{Micro architecture of PE$_k$}\label{fig: pe_archi}
\end{figure}

Based on our S-RCSC decoding algorithm, a corresponding memory efficient decoder architecture is proposed in Fig.~\ref{fig: top_archi}(a), where LMEM, RMEMo, RMEMe, CMEM, SMEM are five LLR memories. LMEM stores the left LLRs in $\mathcal{L}_1,\mathcal{L}_2,\cdots,\mathcal{L}_{n}$. RMEMo stores the right LLRs in $\mathbb{R}^o$, while RMEMe stores the right LLRs in $\mathcal{R}^e_1,\mathcal{R}^e_2, \cdots,\mathcal{R}^e_{n}$. The channel LLR memory, CMEM, stores $\mathcal{L}_{0}$. The soft-output memory, SMEM, stores LLRs in $\mathcal{R}^e_0$.
Our architecture in Fig.~\ref{fig: top_archi}(a) has one processing element array (PEA), which has $P\mbox{ }(P \ll N)$ identical processing elements (PEs). With the concatenation and split method in~\cite{jun_low_mem_list}, LMEM, RMEMo and RMEMe can be implemented with area efficient memories such as register files or SRAMs. Note that each LLR memory in Fig.~\ref{fig: top_archi}(a) is a dual port memory, where each word stores at most $2P$ LLRs. Take the LMEM as an example, the LLRs are stored in this memory in the way shown in Fig.~\ref{fig: top_archi}(b), where $\mathcal{L}_i$ is stored in $\lceil\frac{2^{n-i}}{2P}\rceil$ words and each word stores $2P$ LLRs. LMEM has a total of $W=\sum_{i=1}^n\lceil\frac{2^{n-i}}{2P}\rceil$ words.
The read and write datapaths between LLR memories and the PEA are shown in Fig.~\ref{fig: top_archi}. The read and write address signals and enable signals of each memory are not shown for simplicity. Besides, each memory meeds a bypass buffer~\cite{gross_polar1} to avoid read-write conflict. Suppose each internal LLR (except the channel LLRs) is quantized with $Q$ bits, the width of each read or write data bus of LMEM, RMEMo, RMEMe and SMEM is $T=2PQ$. Suppose each channel LLR is quantized with $Q_c$ bits, then $T_c = 2PQ_c$, where $T_c$ is width of the data buses of the CMEM.
For $k=0,1,\cdots,P-1$, the micro architecture of PE$_j$ is shown in Fig.~\ref{fig: pe_archi}, where the SUM unit outputs the sum of its two input LLRs and the compare-and-select (CAS) unit implements the $f$ function in Eq.~(\ref{equ: f_app}). The write buffer (wB) in Fig.~\ref{fig: pe_archi} is a $Q$-bit register. During each iteration of our S-RCSC decoding algorithm, the left and right LLRs are updated in an interleaved way.


When updating the left LLRs, the computation type control signal, CTC, is set to 0 and the updated left LLRs are written into LMEM. For example, when $\mathcal{L}_i$ needs to be updated, the PEA may need to fetch $\mathcal{L}_{i-1}$, $\mathcal{R}_i^o$ and $\mathcal{R}_i^e$ as shown in Alg.~\ref{algo: left_comp}. When $\mathcal{R}_i^e$ is employed in the updating of $\mathcal{L}_i$, the left LLR selection signal, LLS, is set to 1. Otherwise, LLS = 0. When $\mathcal{R}_i^o$ is needed in order to compute the updated $\mathcal{L}_i$, the control signal, zset, is set to 1. When no right LLRs are needed for the update of $\mathcal{L}_i$, zset = 0. Since the PEA generates at most $P$ updated left LLRs and each word in LMEM stores $2P$ LLRs, $P$ wBs are used to buffer the first $P$ left LLRs, which are written back to LMEM once the next $P$ left LLRs are computed. If $\mathcal{L}_i$ has less than $2P$ LLRs, the left LLR output control signal, LOC, is set to 1.

When updating the right LLRs, CTC = 1, zset = 1 and the updated right LLRs could be written into RMEMo, RMEMe and SMEM. In this case, each PE computes two updated right LLRs in one clock cycle. For our PE architecture in Fig.~\ref{fig: pe_archi}, at most $P$ right LLRs are needed during each clock cycle. However, $2P$ LLRs are read out in parallel for each right LLR memory. In Fig.~\ref{fig: pe_archi}, the segment selection signal, SS, is used to select out $P$ right LLRs from $2P$ inputs.

\subsection{Implementation Results and Comparisons}
\begin{figure} [hbt]
\centering
  \includegraphics[width=2.2in]{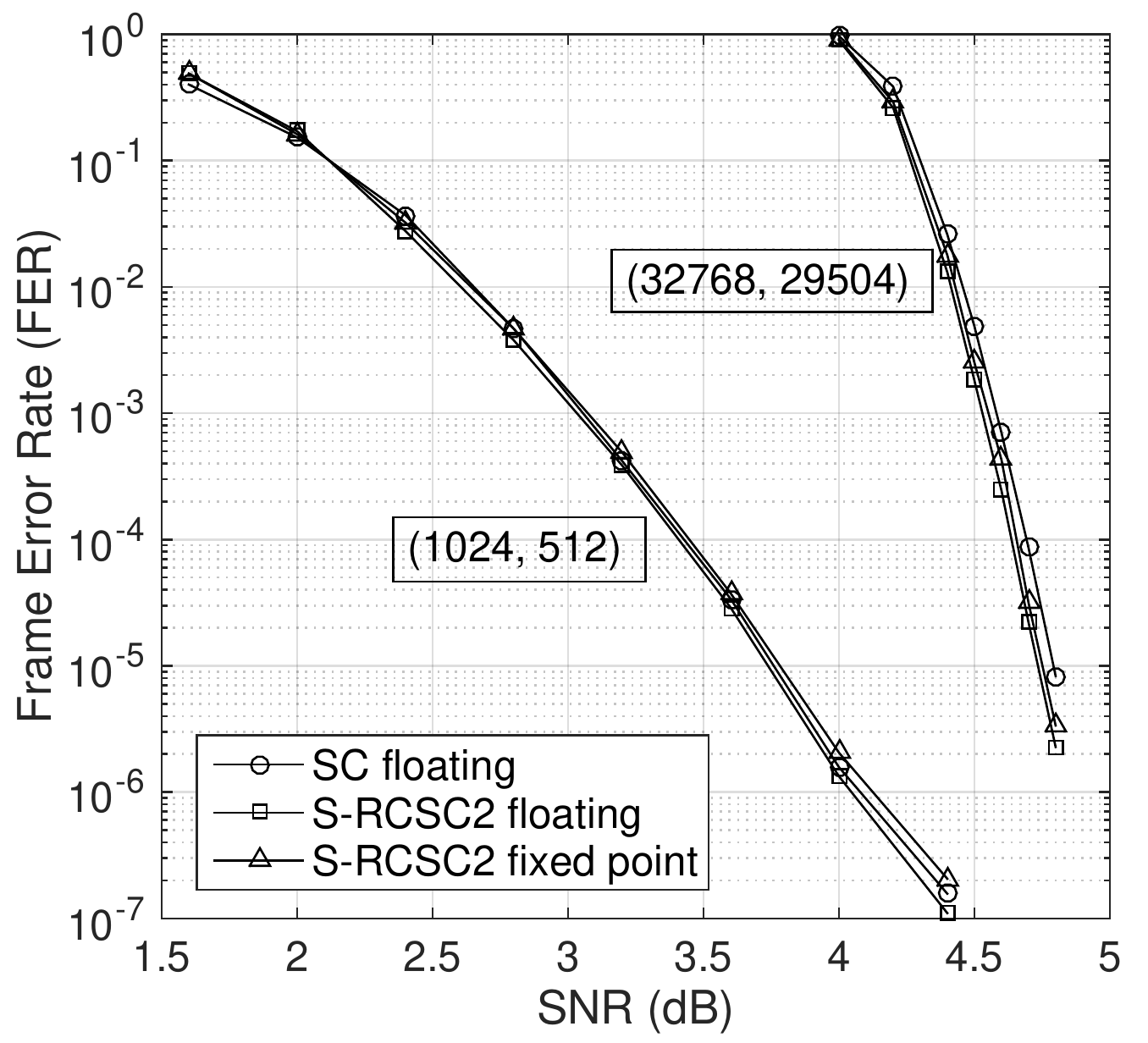}
  \caption{Fixed point error performances}\label{fig: fixed_point}
\end{figure}

The proposed soft-output decoder architecture has been implemented for the (1024, 512) and (32768, 29504) polar codes simulated in this paper. Each channel LLR is quantization with 5 bits, and each of the rest LLR is quantized with 7 bits. The fixed point FER performance of these two decoders are shown in Fig.~\ref{fig: fixed_point}. For our implemented decoders, a maximum of two iterations are allowed. As shown in Fig.~\ref{fig: fixed_point}, our quantization scheme causes negligible error performance degradation. $I_{av}$'s for the two implemented decoders under various SNR are shown in Table~\ref{tab:av_iter}. For both of our decoders, $I_{av}\approx 1$ when FER is below $10^{-3}$.

\begin{table}[!hbt]
  \centering
  \caption{The Average Number of Iterations When $I_M=2$}
  \label{tab:av_iter}
  \footnotesize
  \begin{tabular}{c|c|c|c|c|c|c}
    \hline
    code &\multicolumn{6}{c}{$(1024, 512)$} \\ \hline
     SNR & 1.6 & 2.0 & 2.4 & 2.8 & 4.0 &4.4  \\ \hline
     $I_{av}$ & 1.63 & 1.25 & 1.07 & 1.007 & 1.000003 & 1 \\ \hline\hline
     code & \multicolumn{6}{c}{$(32768, 29504)$} \\ \hline
     SNR & 4.0 & 4.2 & 4.5 & 4.6 & 4.7 &4.8  \\ \hline
     $I_{av}$ & 1.93 & 1.47 & 1.007 & 1.001 & 1.00013 & 1.00001 \\ \hline\hline
  \end{tabular}
\end{table}

Both decoders are synthesized with the Cadence RTL compiler under the TSMC 90nm CMOS technology. The implementation results are shown in Table~\ref{tab:hw_results}, where $N_c$ and CT denote the number of clock cycles per iteration and coded throughput, respectively. Hence, CT = $\frac{Nf}{N_c}$, where $f$ is the achieved frequency. For our decoders with $N=2^{10}$ and $2^{15}$, $P=64$ and 128, respectively. Area efficiency (AE) in Table~\ref{tab:hw_results} denotes the coded throughput normalized by the corresponding area. The implementation results in~\cite{bp_polar_chip_umich} are from the chip fabrication, while the other results are from synthesis. For fair comparisons, our implementation results under 90nm CMOS technology have been scaled to those under 65nm and 45nm.

\begin{table}[!hbt]
  \centering
  \caption{Implementation Results}
  \label{tab:hw_results}
  \footnotesize
  \begin{tabular}{>{\centering}m{1.2cm}|c|c|c|c|c|cm{1cm}}
    \hline
      & \multicolumn{3}{c|}{proposed}&\cite{bp_polar_chip_umich}&\cite{yuan_bp_tsp}&proposed \\ \hline\hline
     $N$ & \multicolumn{5}{c|}{$2^{10}$}& $2^{15}$ \\ \hline
      rate & \multicolumn{5}{c|}{0.5}& 0.9 \\ \hline
     process (nm) & 90 & 65 & 45 & 65 & 45 &90  \\ \hline
     area (mm$^2$) & 0.97 & 0.51* & 0.24\dag & 1.476 & 1.38\ddag & 4.734 \\ \hline
     gate count & \multicolumn{3}{c|}{342k}&-&1.96M&1.677M \\ \hline
     freq. (MHz) &571 & 790*&1142\dag &300 & 500 &518\\ \hline
     $N_c$ &\multicolumn{3}{c|}{610}&10&56&7688\\ \hline
     $I_{av}$&\multicolumn{3}{c|}{1.000066$^\divideontimes$}&6.57$^\sharp$&23$^\|$&1.00001\\ \hline
     CT (Mbps) & 958 & 1326* & 1916\dag &4676 &9000&2208\\ \hline
     AE  & \multirow{2}{*}{987} & \multirow{2}{*}{2600*} &\multirow{2}{*}{7983\dag} & \multirow{2}{*}{3168}&\multirow{2}{*}{6521} &\multirow{2}{*}{466}\\
     (Mb/s/mm$^2$)&&&&&\\ \hline
  \end{tabular}
  \begin{tablenotes}
    \item *Scaled results under the 65nm technology. \dag  Scaled results under the 45nm technology. \ddag Estimated area under the CMOS technology scaling.
    \item $^\divideontimes$The corresponding SNR is 3.5dB. $I_{av}$ will be even smaller for larger SNR. $^\sharp$The corresponding SNR is 4.0dB. $^\|$The corresponding SNR is 3.5dB.
  \end{tablenotes}
\end{table}

As shown in Table~\ref{tab:hw_results}, for the (1024, 512) polar code, the area of our decoder is about 34\% and 17\% of that of the decoder in~\cite{bp_polar_chip_umich} and~\cite{yuan_bp_tsp}, respectively. The coded throughput of our decoder is about 28\% and 21\% of that of the decoder in~\cite{bp_polar_chip_umich} and~\cite{yuan_bp_tsp}, respectively. The area efficiency of our decoder is about 82\% and 122\% of that of the decoder in~\cite{bp_polar_chip_umich} and~\cite{yuan_bp_tsp}, respectively. For the (32768, 29504) polar code, our decoder achieves a coded throughput of 2208Mbps with an area of 4.734mm$^2$ under the TSMC 90nm CMOS technology. Implementation results for such a long code were not shown in~\cite{bp_polar_chip_umich,yuan_bp_tsp}. Compared to the decoder architectures in~\cite{bp_polar_chip_umich,yuan_bp_tsp}, our decoder architecture has advantages in the following aspects.

(a) Compared to the BP decoders~\cite{bp_polar_chip_umich,yuan_bp_tsp}, our decoder has much better error performances when the FER is below $10^{-5}$ as shown in Figs.~\ref{fig: fer1024} and~\ref{fig: fer32k_r0_9} as well as~\cite[Fig. 6]{bp_polar_chip_umich}.


(b) Our decoder architecture is expected to consume less energy on computing updated LLRs compared to the BP decoder.
In this paper, we demonstrate this advantage of our decoder architecture based on an analytical approach.
When the supply voltage is $V$, let $e_{a}^V$ and $e_c^V$ denote the energy used for an addition and a comparison, respectively. Let $E_r$ and $E_b$ denote the average energy used for the decoding of a codeword for the S-RCSC and BP decoders, respectively, where $E_r = I_{av}(N_ae_a^V+N_ce_v^V)$ and $E_b=2Nn(e_a^V+e_c^V)I_{av}^{BP}$. Here, $I_{av}^{BP}$ is the average number of iterations for a BP decoder. $N_a$ and $N_c$ are the numbers of additions and comparisons per iteration for our S-RCSC decoder, respectively.
In this paper, we consider only the energy used for computations over the real field.
When the FER is low, $I_{av}\ll I_{av}^{BP}$ and $I_{av}\approx 1$. Besides, due to our modified message passing schedule and the application of the SSC principle, $N_a<2Nn$ and $N_c<2Nn$. For example, for our (1024, 512) polar code, $N_a=11261$ and $N_c=14332$ while $2Nn=20480$. When SNR is 4.0 dB, $I_{av}=1.000003$ while $I_{av}^{BP}=6.57$ for the decoder in~\cite{bp_polar_chip_umich} as shown in Table~\ref{tab:hw_results}. As a result, $\frac{E_b}{E_r}\approx 8.8$ compared to the BP decoder in~\cite{bp_polar_chip_umich}.

\section{Conclusion}\label{sec:conclusion}
In this paper, we present a more efficient belief propagation decoding algorithm and its hardware architecture. Our decoder architecture shows advantages in terms of error performance and energy efficiency.

\bibliographystyle{IEEEbib}
\bibliography{refs_latest}

\end{document}